\documentclass[12pt]{article}
\usepackage{amsmath,amssymb,epsfig}
\usepackage{cite}

\usepackage{color}
\input{colordvi.tex}
\def\unit{{\relax{\rm 1\kern-.26em I}}}

\def\nn{\nonumber\\}
\newdimen\Tdim
\def\ispan{{\setbox0=\hbox{i}%
\Tdim\ht0\advance\Tdim\dp0\rule[-\dp0]{0pt}{\Tdim}}}
\def\jspan{{\setbox0=\hbox{j}%
\Tdim\ht0\advance\Tdim\dp0\rule[-\dp0]{0pt}{\Tdim}}}
\def\Tspan#1{{\setbox0=\hbox{#1}%
\Tdim\ht0\advance\Tdim\dp0\advance\Tdim.55ex\rule[-\dp0]{0pt}{\Tdim}\box0}}

\makeatletter

\renewcommand\section{\@startsection {section}{1}{\z@}%
                                  {-3.5ex \@plus -1ex \@minus -.2ex}%
                                  {2.3ex \@plus.2ex}%
                                  {\normalfont\large\bfseries}}

\renewcommand\subsection{\@startsection{subsection}{2}{\z@}%
                                    {-3.25ex\@plus -1ex \@minus -.2ex}%
                                    {1.5ex \@plus .2ex}%
                                    {\normalfont\normalsize\bfseries}}

\newcount\hour \newcount\minute
\hour=\time \divide \hour by 60
\minute=\time
\def\now{%
\ifnum \hour<13
 \ifnum \hour=0 \advance \hour by 12 \number\hour:\else \number\hour:\fi%
    \ifnum \minute<10 0\fi%
    \number\minute%
\ A.M.%
\else \advance \hour by -12 \number\hour:%
 \ifnum \minute<10 0\fi%
 \number\minute%
 \ P.M.%
\fi%
}

\makeatother

\begin{document}

\baselineskip=18pt  
\numberwithin{equation}{section}  
\allowdisplaybreaks  



%
%


\thispagestyle{empty}

\vspace*{-2cm}
\begin{flushright}
\end{flushright}

\begin{flushright}
KUNS-2437\\
DESY 13-033 \\
OCU-PHYS-381
\end{flushright}

\begin{center}

\vspace{1.4cm}

\vspace{0.5cm}
{\bf \Large Cosmic R-string in thermal history}
\vspace*{0.5cm}

\vspace{0.5cm}

{\bf
Kohei Kamada$^{1}$, Tatsuo
Kobayashi$^{2}$, Keisuke Ohashi$^{3}$ and Yutaka Ookouchi$^{2,4}$}
\vspace*{0.5cm}

\vspace*{0.5cm}

$^1$ {\it Deutsches Elektronen-Synchrotron DESY,
Notkestrasse 85, D-22607 Hamburg, Germany }\\

\vspace{0.1cm}

$^{2}${\it Department of Physics, Kyoto University, Kyoto 606-8502, Japan}\\

\vspace{0.1cm}

$^{3}${\it Department of Mathematics and Physics, Osaka City
 University, Osaka 558-8585, Japan  }\\

\vspace{0.1cm}

$^{4}${\it The Hakubi Center for Advanced Research, Kyoto University, Kyoto 606-8302, Japan }\\

\vspace*{0.5cm}

\end{center}

\vspace{1cm} \centerline{\bf Abstract} \vspace*{0.5cm}

We study stabilization of an unstable cosmic string associated with spontaneously broken $U(1)_R$ symmetry, which otherwise causes a dangerous roll-over process. We demonstrate that in a gauge mediation model, messengers can receive enough corrections from the thermal plasma of the supersymmetric standard model particles to stabilize the unstable modes of the string.

\newpage
\setcounter{page}{1} 



\section{Introduction}

The string landscape is a fascinating idea to reveal the nature of the Universe and 
suggests us that the vacuum structure of a field theory itself may be complicated. In phenomenological model building, this idea gives us a good opportunity to revisit supersymmetry (SUSY) breaking and its phenomenological applications\cite{Intriligator:2006dd,OO1,KOO,Abe:2007ax} (See \cite{rev1,rev2,rev3} for reviews and references therein). 
In particular, the idea that we live in a metastable SUSY-breaking vacuum is now one of the most promising 
scenarios from both phenomenological and theoretical viewpoints. As is
emphasized recently in \cite{our1}, if such a landscape of vacua is
realized in nature, the existence of a certain type of solitonic objects can have an important meaning. Such a soliton can be viewed as an energetic impurity which causes semiclassical vacuum decay via rolling-over the potential hill toward a lower energy vacuum. This idea was proposed almost thirty years ago by several authors \cite{Steinhardt:1981ec,Hosotani,Yajnik:1986tg}. Here we revisit it in light of the landscape of vacua and study how to avoid the roll-over problem
in realistic SUSY-breaking models. 

In this paper, as an illustration, we consider a gauge mediation model
with spontaneously broken $U(1)_R$ symmetry and supersymmetry. So far,
various attempts have been done to build several types of models of
spontaneously broken R-symmetry
\cite{Kitano1,Kitano2,Shih:2007av,Ferretti:2007ec,Cho:2007yn,Abel:2007jx,Aldrovandi:2008sc,Carpenter:2008wi,Giveon:2008ne,Sun:2008va,Azeyanagi:2012pc,R1,R2,Stone}. One
of the lessons worth emphasizing is a connection between metastability
and large gaugino masses. To generate large gaugino masses in gauge
mediation models, as is nicely formulated in \cite{KS} (see
\cite{Ookouchi} for a review), one needs a tachyonic direction that
leads to the SUSY vacuum in the pseudo-moduli space if the low energy effective theory is approximated by a renormalizable generalized O'Raifeartiagh model. In the models with spontaneously broken R-symmetry, the tachyonic direction exists at the origin \cite{Shih:2007av}. 
Since the $U(1)_R$ symmetry is spontaneously broken at the
SUSY-breaking vacuum, a global cosmic string can be formed by the
Kibble-Zurek mechanism\cite{Kibble:1980mv,Zurek} at some time
in the cosmic history\footnote{In \cite{Eto:2006yv} solitons in
  metastable SUSY breaking vacuum were investigated and later, such
  solitons were used for cosmological constraints on gauge mediation
  models\cite{Hanaki}.
In addition, a metastable system with U(1) symmetry would 
decay through Q-balls. 
That leads to another cosmological constraint \cite{Barnard:2010wk}. }. 
In the core of usual strings, the energy is
large and the symmetry is restored. However, as mentioned above, at a
symmetry restoring point in this model, there exists a tachyonic direction along the messenger direction. Therefore, the core of the string in this case slides down to a lower vacuum and the string transforms into the metastable tube-like soliton which we call R-tube. In \cite{our1}, we investigated the various aspects of the R-tube and found that there can be light unstable modes. In these reasons, a gauge mediation model with spontaneously broken R-symmetry is an ideal example to demonstrate implications of such complicated vacua in a realistic situation (see \cite{Kumar1} for relevant earlier works). 

In this paper, we claim that such unstable modes of cosmic R-string/R-tube can be stabilized by the thermal potential generated by the thermal plasma of the supersymmetric standard model particles even with sufficiently small reheating temperature. 
One may think that  
as the universe expands, the temperature decreases and eventually the thermal potential becomes too small to stabilize the unstable modes. However, since R-symmetry is explicitly broken by the gravity effect, axionic domain walls connecting the R-strings are formed when the Hubble parameter becomes comparable to the R-axion mass and finally all string network disappears. Therefore, if the thermal protection of the unstable modes is valid until the time of their decay, it is plausible to conclude that such models are free from the disastrous roll-over problem. 

The paper is organized as follows. In section 2, we set up  a gauge mediation model with spontaneously broken $U(1)_R$ symmetry. We show similarity between global string solution in \cite{our1} and that of the present model. We review some aspects of the string solution and add some new comments on the solution. 
In section 3, we demonstrate that the thermal potential generated by the
thermal plasma of standard model particles lifts the potential of
messengers and stabilized the unstable modes of R-string/R-tube. In
section 4, we consider the model in the inflaton oscillation dominated era
of the expanding universe and study the vacuum selection\footnote{See \cite{selection1,selection2,selection3,selection4,selection5} for early attempts for vacuum selection by exploiting the thermal potential.} and stabilization of strings. In section 5, we comment on relevant issue and discussions. In appendix A and B, we show some technical details on R-string solution and tachyonic mass around it.

\section{Model with spontaneously broken R-symmetry}

\subsection{Setup of Model}\label{sec:setup}

Let us promote the original model studied in \cite{our1} to more
realistic gauge mediation model. We introduce the superfields, $X$, $\phi$ and $\tilde{\phi}$.
A pair of $\phi$ and $\tilde{\phi}$ correspond to 
the messenger fields, which have the vector-like representation, 
$\bf R \oplus \bf \bar{R}$,
under the standard gauge group, $SU(3) \times SU(2) \times U(1)$.
For simple illustration, here we restrict ourselves to 
the $U(1)$ gauge part of $SU(3) \times SU(2) \times U(1)$ as the 
visible sector.
Then, we consider the messenger fields, $\phi$ and $\tilde{\phi}$, 
which are singlets and have the $U(1)$ charges 1 and $-1$.
Its extension to the full $SU(3) \times SU(2) \times U(1)$ gauge 
theory is straightforward as we will give a comment later.

We assign R-charges to these superfields as, $R[X]=2$ and
$R[\phi]+R[\tilde{\phi}]=0$, and the superpotential is obtained by 
\begin{eqnarray}
W=\kappa X\phi \tilde{\phi} - \mu^2 X.
\end{eqnarray}
Note that we introduced $\kappa$ to control the vacuum selection. As
is used in \cite{NakaiOokouchi,FKT} by taking $\kappa$ to be
small, the tachyonic mass $|m_\phi|$ along $\phi$, $\tilde{\phi}$ can be small,
which is favorable for the realistic vacuum selection. 
We consider the effective K\"ahler metric, 
\begin{equation}
g_{X\bar{X}}^{-1}=1-{1\over 2m^2} |X|^2 +{\lambda \over 4 m^4}|X|^4,
\quad g_{\phi \bar{\phi}}^{-1}= g_{\tilde{\phi} \bar{\tilde{\phi}}}^{-1}=1,\quad
g^{-1}_{X\bar{\phi}}=g^{-1}_{\phi \bar{X}}=g^{-1}_{X\bar{\tilde{\phi}}}=g^{-1}_{\tilde{\phi} \bar{X}}=0, 
\end{equation}
which are similar to those in \cite{our1}.
We take all of parameters, $\kappa$, $\mu^2$, $m^2$ and $\lambda$, 
to be real and positive.
This model possesses the SUSY vacuum with the moduli space, 
\begin{eqnarray}
X=0, \qquad \phi \tilde{\phi}=  {\mu^2 \over \kappa },
\end{eqnarray}
and the SUSY breaking vacuum,
\begin{eqnarray}
X= X_0\equiv \frac{m}{\sqrt{\lambda}}, \qquad \phi =0,
\end{eqnarray}
where the $U(1)_R$ symmetry is also broken.
The former is the true vacuum, while the latter is 
the metastable vacuum whose vacuum energy is 
$V=\mu^4\left(1-\frac{1}{4\lambda}\right)$.

For later convenience, let us introduce dimensionless variables by
\begin{eqnarray}
X = \frac{m}{\sqrt{\lambda}}{A} ,\quad ,\tilde{\phi} = {\mu\over \sqrt{\kappa}} \tilde{s},\quad \phi={\mu\over \sqrt{\kappa}} s, \quad  x_\mu = \frac{m}{\sqrt{\lambda}\mu^2}\tilde x_\mu,
\quad \epsilon = \frac{\sqrt{2}\sqrt{\lambda} \mu}{\sqrt{\kappa}m},
\label{eq:dimmless}
\end{eqnarray}
then the Lagrangian is of the form
\begin{eqnarray}
{\cal L} &=& \mu^4\left[
\frac1{{\cal V}({A})}
\left|\tilde \partial_\mu {A}\right|^2
+ {1\over 2}\epsilon^2\left|\tilde \partial_\mu s\right|^2+ {1\over 2}\epsilon^2\left|\tilde \partial_\mu \tilde{s}\right|^2 - {\cal V}({A})| s\tilde{s}-1|^2
- \frac{2}{\epsilon^2}|{A}|^2 (|s|^2+|\tilde{s}|^2) \right. \nonumber \\ 
&& \left. -{g^2 \over
    8\kappa^2}(|s|^2-|\tilde{s}|^2)^2
\right] + \cdots,
\label{eq:model_sol3}
\end{eqnarray}
where the ellipsis denotes the fermionic partners and 
the supersymmetric standard model particles including gauge fields and 
matter fields, and 
${{\cal
    V}(A)}$ is given by 
\begin{eqnarray}
 {{\cal V}(A)}=1-\frac1{4\lambda}+\frac1{4\lambda}(1-|A|^2)^2.
\end{eqnarray}
Basic ideas we will show below can be simply demonstrated by this simplified model. 
Since throughout this paper, we do not consider a soliton with the
winding number of $U(1)$, we can take the vacuum expectation value (VEV)
 of gauge fields vanishing in constructing solitons.

Various quantities appeared in the first line of (\ref{eq:model_sol3}) are characterized by two dimensionless parameters
$\lambda$ and $\epsilon$.
For instance, the existence of the SUSY breaking vacuum requires
\begin{eqnarray}
 1> \frac1{4\lambda}>0,\quad   \frac{2}{\epsilon^2}+\frac1{4\lambda}>1.
\label{eq:VacuumCond}
\end{eqnarray} 
Later, we consider the vacuum selection in the early stage of the Universe. Since in the early Universe field values are assumed  to be around the
origin, if the tachyonic mass of $X$, $|m_X|$, is larger than one of
the messenger field, $|m_\phi|$, 
one may expect that the supersymmetry breaking vacuum is preferable. 
In the dimensionful representation, these two tachyonic masses are given by
\begin{equation}
m_X^2=-{\mu^{ 4}\over 2m^2},\quad m_{\phi}^2=-\kappa \mu^2.
\end{equation}
Thus, the following inequality
\begin{eqnarray}
 \frac{m_X^2}{m_\phi^2}>1\quad  \Leftrightarrow \quad \frac2{\epsilon^2}<\frac1{2\lambda} , \label{redline}
\end{eqnarray}
is required for selecting the SUSY breaking vacuum. Note that, according to the reparametrization (\ref{eq:dimmless}),
a small $\kappa$ with fixing $m, \mu$ corresponds to a large $\epsilon$.
As discussed later in section \ref{sec:Universe}, the thermal effect  
turns out to loosen this condition.

For later convenience, we show the gravitino and axion masses,
\begin{equation}
m_{3/2}\simeq \sqrt{g^{-1}(X_0)} {\mu^2\over \sqrt{3} M_{\rm pl}}, \quad
m_a=3^{3/4}m_{3/2}(g^{-1}(X_0))^{-1/4} \sqrt{2M_{\rm pl}\over X_0} ,
\end{equation}
where $g^{-1}(X_0)=1-{1\over 4\lambda}$ and $M_{\rm pl}$ denotes 
the reduced Planck mass. Using the  gravitino mass $m_{3/2}$ 
with  dimensionless parameters $\lambda,\epsilon,\kappa$,  many scales
can be rewritten as, for instance,
\begin{eqnarray}
 \left(\frac{m_a^2}{m_X^2}\right)^2\simeq \frac{48
  \lambda^{3/2}\sqrt{12\lambda-3}}{\epsilon^2\kappa} \times
  \frac{m_{3/2}}{M_{\rm pl}} \simeq 
2.1\times 10^{-18}\times  \frac{2}{\epsilon^2} 
\frac{(4\lambda)^{3/2}\sqrt{4\lambda-1}}{\kappa} \times
\frac{m_{3/2}}{\rm GeV}. \label{eq:ratio}
\end{eqnarray}
In terms of  \eqref{eq:VacuumCond} and the scale of the gravitino mass in gauge mediation, we observe $|m_X|\gg m_a$ in wide range of parameter space $(\lambda , \kappa)$. Therefore, we assume the condition in this paper.

\subsection{R-string review }

Here we quickly review the R-string studied in \cite{our1}. 
Since the R-string solution satisfies the D-flatness condition, the
equation of motion for R-string becomes exactly the same as the one
studied in \cite{our1}. 
Hence, we simply see some of aspects.

\begin{figure}[htbp]
\begin{center}
 \includegraphics[width=0.47\linewidth]{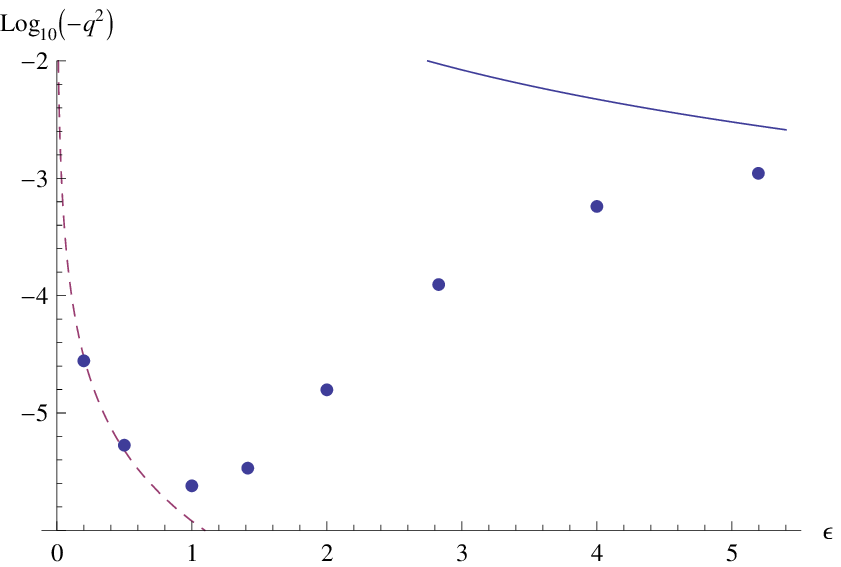}\quad
  \includegraphics[width=.47\linewidth]{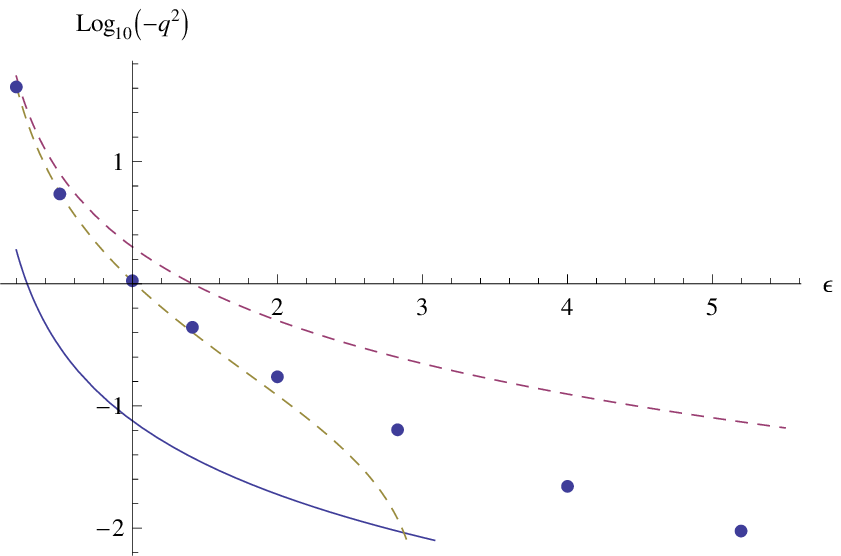}
\vspace{-.1cm}
\caption{\sl Tachyonic masses for unstable modes of $n=1$ and $n=2$ R-strings with $\lambda=0.27$. 
Solid lines represent $\epsilon^2 q^2=-2(1-1/\sqrt{4\lambda})$ and a dashed line in the left panel is 
$\displaystyle \epsilon^2 q^2=-2\times
 6\times 10^{-7}$ and 
dashed lines in the right panel are  $\displaystyle \epsilon^2 q^2=-2,
 -2\left(1-0.47
 \epsilon^{2/3}\right)$.} 
\label{TachyonMass1}
\end{center}
\end{figure}

\begin{figure}[htbp]
\begin{center}
 \includegraphics[width=0.47\linewidth]{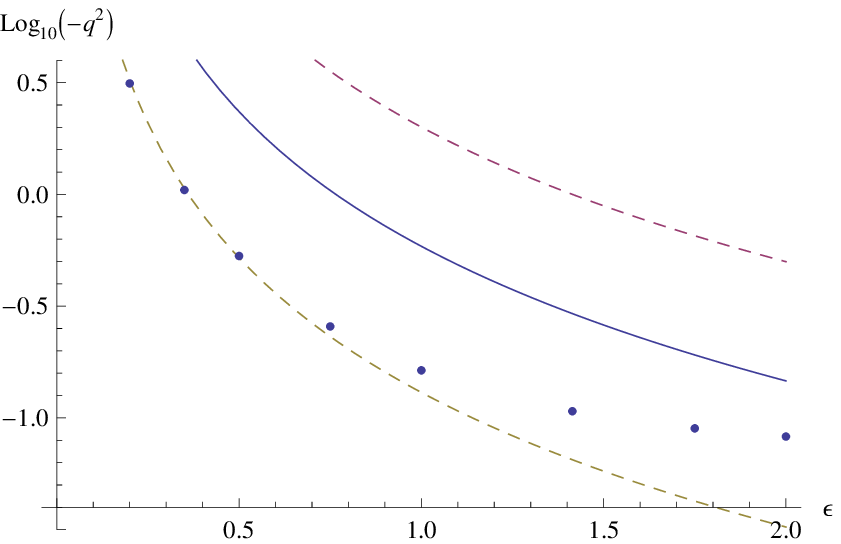}
  \includegraphics[width=.47\linewidth]{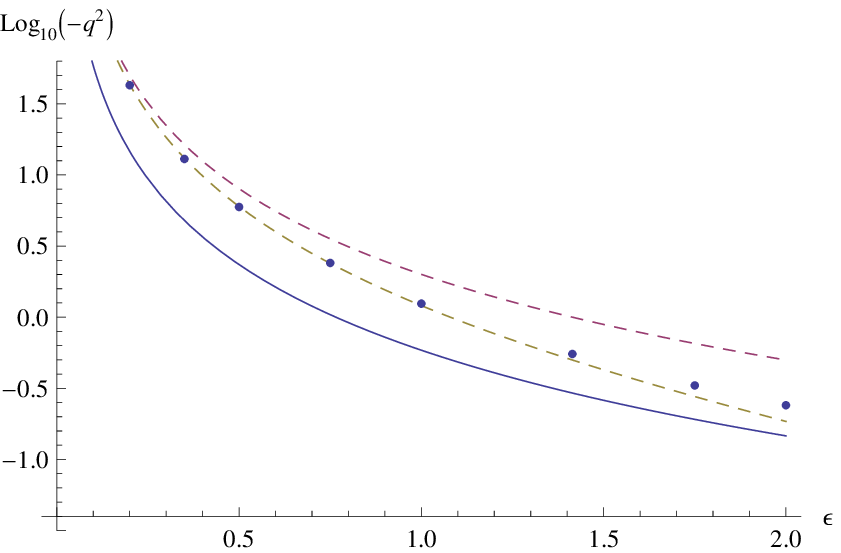}
\vspace{-.1cm}
\caption{\sl Tachyonic masses for unstable modes of $n=1$ and $n=2$ R-strings with $\lambda=1/2$. Dashed lines in the left
 panel are 
 $\displaystyle \epsilon^2 q^2=-2,-2\times 0.065$ and 
dashed lines in the right panel are  $\displaystyle \epsilon^2 q^2=-2,
 -2\left(1-0.39\epsilon^{2/3} \right)$. }
\label{TachyonMass2}
\end{center}
\end{figure}

The R-string solution without a hole inside, $s=\tilde{s}=0,$ is a
solution of our model. The R-string solution 
$A(\tilde x^\mu)=A_n^{\rm sol}(\rho)
e^{in\theta},(\rho e^{in\theta}=\tilde x_1+i\tilde x_2, A_n^{\rm sol}\in
\mathbb R_{\ge 0})$ with a given winding number $n$ 
is defined by imposing the equation of motion derived from 
the following `reduced' action,
\begin{eqnarray}
\tilde S=-2\pi \int_0^\infty d\rho \rho 
\left\{
\frac{1}{{\cal V}(A)} 
\left(\left(\frac{d A}{d
       \rho}\right)^2+\frac{n^2}{\rho^2}A^2\right)+{\cal V}(A)\right\},
\end{eqnarray}
with boundary conditions $A(0)=0$ and  $A(\infty)=1$. The R-string
solution corresponds to the solution of the system described by the above Lagrangian 
and we can find it numerically (see appendix \ref{propertyan} for
details of the solution.). However, as shown in \cite{our1} such an
R-string is unstable and transforms into an R-tube with non-zero
$s=\tilde{s}$ inside. Since our main motivation is to stabilize such
unstable modes, let us quickly review the analysis to get ``masses'' of the modes. 
To see that, let us consider a linearized equation for $s$ around the R-string solution, 
\begin{eqnarray}
 -\frac1\rho \frac{d}{d\rho}\left(\rho \frac{d s}{d\rho }\right)+
V_{\rm pot}(A_n^{\rm  sol}(\rho)) \,s= q^2_n s,\qquad 
V_{\rm pot}(A)\equiv -\frac{2}{\epsilon^2}{\cal
			 V}(A)+\frac{4}{\epsilon^4}|A|^2. 
\label{eq:tachyonic_mass}
\end{eqnarray}
Here, the eigenvalue  $q^2_n$ for the R-string solution with the winding
number $n$ depends on $\lambda$ and $\epsilon$, i.e.
$q^2_n=q^2_n(\lambda,\epsilon)$, and,
an observation $0\le A_n^{\rm sol}\le 1$  tells us the lower bounds of
$V_{\rm pot}$ and $q^2_n$ as 
\begin{eqnarray}
 V_{\rm pot}(A)\ge V_{\rm pot}(0)=-\frac{2}{\epsilon^2},
\quad \Rightarrow    \quad q_n^2>-\frac{2}{\epsilon^2}.
\end{eqnarray}
Taking 
the limits of $\lambda \to \infty$ or $\epsilon\to 0$, we
can show that 
\begin{eqnarray}
\lim_{n\to \infty}\epsilon^2 q_n^2=-2,\quad \lim_{\lambda \to \infty} \epsilon^2 q_n^2=-2,\quad 
\lim_{\epsilon\to 0} \epsilon^2 q_{n>1}^2=-2.
\end{eqnarray} 
For the proof of these facts,  see Appendix.\ref{sec:tachyon}. 
Furthermore,
we experientially find that, with several typical sets of parameters 
$\lambda$ and $\epsilon$,
\begin{eqnarray}
 0< -\epsilon^2q^2_1 < -\epsilon^2q^2_2<\cdots < 2,   
\end{eqnarray}
and interestingly  
\begin{eqnarray}
 -\epsilon^2q^2_1<2\left(1-\frac1{2\sqrt{\lambda}}\right)<-\epsilon^2q^2_2,
\label{eq:funnyineq}
\end{eqnarray}
although we have no proof. 
We numerically solved this equation for several parameters as shown in Figure \ref{TachyonMass1} and \ref{TachyonMass2}. 
The left and right panels in both figures correspond to 
$n=1$ and 2, respectively.

For later convenience, we introduce the ``physical'' mass 
for the unstable mode.  The eigenvalue $q_n^2$ for the R-string with winding number $n$ 
is translated in terms of the mass for the canonically normalized messenger field 
as
\begin{equation}
m_{n}^2=q^2_n  \left({\sqrt{\lambda} \mu^2\over m } \right)^2=
 \frac{\epsilon^2q_n^2}{2} \times \kappa \mu^2.   
\end{equation}
We will use it later for the physical considerations. Here an non-trivial fact $m_\infty^2=m_\phi^2=-\kappa \mu^2$ is observed.

\section{R-string in thermal plasma}

Now we turn to the stabilization of the R-strings. 
As shown in the previous section, R-strings in the (Minkowski background) vacuum 
are unstable and unavoidably deform to R-tubes. 
As discussed in \cite{our1}, R-tubes are basically unstable objects and 
cause, in turn, the roll-over process to push the whole Universe into the unwanted SUSY vacuum with a finite time. 
Therefore, one may consider that the SUSY-breaking models with the spontaneous R-symmetry breaking 
are disfavored if the R-symmetry has once been restored in the cosmic history and R-strings have been formed.

However, the realistic Universe is neither described by the Minkowski space time nor in the vacuum. It contains 
several ingredients that can deform the potential for the R-string sector. 
It may be possible to stabilize the R-strings. The most effective contribution would be the thermal potential 
in the existence of the thermal plasma and hence we focus on the thermal effect in the following\footnote{The 
Hubble induced masses generated from the Planck suppressed interaction between the R-string sector and 
inflaton also exist during the inflaton oscillation dominated era before reheating. However, 
we can show easily that it cannot stabilize R-strings sufficiently in the wide range of parameter space.   }. 

Here we should note that we do not have to stabilize R-strings up to the present time. 
As explained in \cite{OurII}, R-string networks, if they are stabilized, vanish when the Hubble parameter 
of the Universe becomes sufficiently small, $H\simeq H_a\equiv m_a$ \cite{Sikivie:1982qv}.  
This is because the constant term is added 
in the supergravity superpotential to compensate the vanishing
cosmological constant, and such a constant term 
explicitly breaks the R-symmetry and 
makes the R-string healthily unstable. From the explicit R-symmetry
breaking effect, the R-axion, which is the 
phase direction of the SUSY-breaking field acquires a mass. Thus, when the Hubble parameter decreases sufficiently, 
the R-string network turns to the R-string-domain wall system, which collapses to R-axion particles immediately 
\cite{Sikivie:1982qv}. 
Therefore, we only have to stabilize R-strings up to 
the Hubble time corresponding to $H=H_a$.  

We also note that high reheating temperature is not necessarily required in our discussion. 
In the gauge mediated SUSY-breaking, there is a severe constraint on the reheating temperature 
from the gravitino problem\footnote{For the cosmological constraint from R-axions, see \cite{OurII}.}. 
Therefore, thermal plasma seems to be difficult to modify the potential for the R-string sector. 
However, even before reheating during the inflaton oscillation dominated era, 
thermal plasma of supersymmetric standard model particles 
{\it does exist} from the partial decay of inflaton quanta, of which temperature is larger than 
the  reheating temperature. Thus, it is plausible to consider the thermal potential for the R-string sector 
that is generated by the thermal plasma. In this section, we demonstrate that the unstable modes shown in 
the previous section can be stabilized by the thermal effect assuming the existence of thermal plasma 
with sufficiently high temperature. 
We will revisit the cosmic history with R-strings in the next section.

\subsection{Thermal potential for messenger}

In this subsection, we study the thermal potential in the hidden sector
generated by the supersymmetric standard model thermal plasma through
the gauge interaction. 
Here we assume that the temperature of thermal plasma is relatively low, $T<\kappa X_0$ and 
the moduli field and messenger field are not thermalized\footnote{For smaller temperature, the number density of  
heavy fields is Boltzmann suppressed, if ever, and hence they cannot be in thermal equilibrium in the expanding Universe. 
The mass of messenger fields around the R-string core is typically small enough to be thermalized. 
However, the width of R-string is too small for the messenger fields to be thermalized and 
we do not consider the effect of messenger thermalization. }
since we are interested in the last stage when the stabilization mechanism is effective. 

Since nonvanishing messenger field values generate the Standard model gauge boson (and gaugino) mass, 
their one-loop effective thermal potential generates thermal correction to the scalar potential for the  messenger 
fields (see \cite{Quiros} for a review and references therein), 
\begin{eqnarray}\label{eq:thermal-V}
V_{\rm thermal}(\phi,{\tilde \phi})\simeq 3{T^4 \over 2\pi^2}J_B(m_B^2/T^2),  \quad m_B^2 ={g^2 \over 2} \left( |\phi|^2+|\tilde{\phi}|^2 \right),\label{eq:thermal}
\end{eqnarray}
where factor $3$ comes from the degree of freedom of the massive
$U(1)$ vector bosons\footnote{Remember that for simplicity we consider 
only the singlet messenger fields $\phi$ and $\tilde \phi$ with 
the $U(1)$ charges $\pm 1$.
It is straightforward to proceed the argument to the full $SU(3)
\times SU(2) \times U(1)$ theory.} and
\begin{equation}
J_B(y)=\int_0^{\infty} dx \ x^2 \, \log \left[ 1-e^{-\sqrt{x^2+y }} \right]. 
\end{equation}
For smaller messenger field values, we can use the high temperature expansion, 
\begin{eqnarray}
&&J_B (m_B^2/T^2)=-{\pi^4 \over 45} +{\pi^2 \over 12}{m_B^2 \over T^2 } -{\pi \over 6}\left( {m_B^2 \over T^2} \right)^{3/2}  -{1\over 32}{m_B^4 \over T^4}\log {m_B^2 \over a_B T^2} \nonumber \\
&&{}\qquad \qquad \qquad \qquad \qquad -2\pi^{7/2} \sum_{l=1}^{\infty} (-1)^l {\zeta(2l+1)\over (l+1)!} \Gamma \left( l+{1\over 2} \right) \left({m_B^2 \over 4\pi^2 T^2} \right)^{l+2}, \label{JB}
\end{eqnarray}
with $\log a_B=5.4076$. 
Then, the messenger fields acquire the effective mass, $m_\phi^T$,
as
\begin{eqnarray}
V_{\rm thermal} (\phi, {\tilde \phi}) \simeq
 \frac{g^2}{16}T^2(|\phi|^2+|{\tilde \phi}|^2),\quad
 \Rightarrow\quad m_\phi^T=\frac{g}4 T. \label{eq:thermal-V-2}
\end{eqnarray}
We can expect that the R-string is stabilized if the thermal mass is larger than 
the physical mass of the tachyonic mode discussed in the previous
section. 
The ratio of the messenger mass to the tachyonic mode mass is given by
\begin{equation}
{\kappa X_0 \over |m_n|}={2\over \epsilon^2|q_n|},
\end{equation}
which is larger than one  in a wide range of parameter space. 
Thus, even in a low temperature, $T<\kappa X_0$, the thermal mass can be comparable to the 
physical tachyonic mode mass for the R-string, and hence it is enough to use the above thermal potential to study stability of tachyonic modes. 

To complete the argument on thermal effects, we comment on the higher temperature situation, $T>\kappa X_0$. 
In this case, the messenger fields are also thermalized and extra contribution to thermal effective potential to 
messenger direction is generated,
\begin{equation}
V\simeq 4{T^4 \over 2\pi^2}(J_B(m_B^2/T^2)  -J_F(m_B^2/T^2)) , \quad m_B^2= {g^2 \over 2} \left( |\phi|^2+|\tilde{\phi}|^2 \right)
\end{equation}
where 
\begin{equation}
J_F(y)=\int_0^{\infty} dx \ x^2 \, \log \left[ 1+e^{-\sqrt{x^2+y }} \right]. 
\end{equation}
Here, the fermion contribution comes from the diagram with gaugino and messengino, 
whereas the additional scalar contribution comes from the messenger scalar
loop, which acquires mass from the D-term. 
We should note that the thermal potential for $X$ is also generated by messenger fields, $V\sim \kappa^2  T^2 |X|^2+\cdots$. 
However, if $\kappa$ is small, the thermal potential can be much smaller than the zero-temperature mass of $X$ at
$T \gtrsim \kappa X_0$. 
Thus, we can neglect thermal corrections to the potential for the $X$ field here. 

Finally, we comment on the extension of the gauge theory. 
So far, we have considered the singlet messenger fields, $\phi$ and 
$\tilde \phi$, with the $U(1)$ charges $\pm 1$ for simplicity.
It is straightforward to extend our discussions for 
generic messenger fields, $\phi$ and 
$\tilde \phi$ with the vector-like representations, 
${\bf R} \oplus \bf{\bar R}$, under the $SU(3)\times SU(2) \times
U(1)$ gauge symmetry.
In such a generic case, all of the gauge bosons contribute 
to induce the thermal potential of the messenger fields.
Such a thermal potential is obtained by replacing $g^2$ 
in (\ref{eq:thermal-V}) 
and (\ref{eq:thermal-V-2}) by 
\begin{equation}
g^2_{eff}= g^2_3 C_2^{(3)}({\bf R}) + g^2_2 C_2^{(2)}({\bf R}) + 
g^2 q^2_\phi,
\end{equation}
where $C_2^{(3)}({\bf R})$ and $C_2^{(2)}({\bf R})$ denote the 
quadratic Casimir indices of the ${\bf R}$ representations of 
$SU(3)$ and  $SU(2)$, respectively, and $q_{\phi}$ denotes 
the hypercharge of $\phi$.
Thus, our discussions in the following sections can be 
extended into the full visible 
$SU(3)\times SU(2) \times U(1)$ gauge theory by replacing 
$g^2 \rightarrow g^2_{eff}$.

\subsection{Tachyonic mode around R-string}

Now we are ready to study thermal effects on R-string. Firstly, let us
see how the tachyonic mass of the R-string can be modified by the thermal
effect in (\ref{eq:thermal}). 
To study this, we have only to pay attention to 
an infinitesimal fluctuation $|\phi|^2=|\tilde \phi|^2=\mu^2 |s|^2
/\kappa \ll
\mu^2/\kappa$, which indicates that 
we can use the high temperature approximation,
\begin{eqnarray}
V_{\rm thermal}=\mu^4\left(-\frac{\pi^2}{30}{T^4\over \mu^4}  
+\frac{g^2 }{8\kappa} {T^2 \over \mu^2} |s|^2+\cdots\right).  
\end{eqnarray}
The second term in the above gives a constant shift of the potential 
in (\ref{eq:tachyonic_mass}) as
\begin{eqnarray}
 V_{\rm pot}(A) \quad \to \quad V_{\rm pot}(A,T)=V_{\rm pot}(A) 
+\frac{g^2 }{8 \kappa \epsilon^2} {T^2 \over \mu^2},
\end{eqnarray}
and resultantly  a mass eigenvalue $q^2_n(T)$ also accepts the constant shift,
\begin{eqnarray}
 q^2_n(T)=q^2_n+\frac{g^2 }{8 \kappa \epsilon^2} {T^2 \over \mu^2}.
\end{eqnarray}
It is straightforward to define the $n$-th critical temperature, $T_n$, where 
 $q^2_n(T_n)\equiv 0$, that is,
\begin{eqnarray}
 T_n=\frac{4 }{g}\times \sqrt{-\frac{\epsilon^2 q_n^2}2 \times \kappa
  \mu^2}= \frac{4}{g} |m_n|,
\end{eqnarray}
and then we find the following properties 
\begin{eqnarray}
 0<T_1<T_2<\cdots <\frac{4\sqrt{\kappa}}{g} \mu\equiv T_{\infty},\qquad 
 \lim_{\lambda \to \infty}T_n=\frac{4\sqrt{\kappa}}g \mu,\quad
\lim_{\epsilon\to 0}T_{n>1}=\frac{4 \sqrt{\kappa}}g \mu.
\end{eqnarray}
At a temperature $T$ larger than $T_n$, the tachyonic mass changes to
massive one $q_n^2(T)>0$ and therefore the R-string with a winding number
$n$ is stable.

As an illustration, we show the low energy effective potential for the light mode. As in \cite{our1}, to uncover the existence of light unstable mode, which is almost frozen in the relaxation method, we vary the initial profile function.  To be concrete, the following initial conditions with various values of $\rho_0$ are used for $A=f(\rho)e^{in \theta},s=h(\rho)$
\begin{eqnarray}
&& f(\rho)=\frac{1+\tanh(2(\rho-\rho_0))}2 \tanh(\rho), \\ \nonumber 
 && h(\rho)=\frac{1-\tanh(2(\rho-\rho_0))}2
\frac{1+\tanh(2(\rho+\rho_0))}2.
\end{eqnarray}  
Also, to represent the size of the tube, we define the following parameter, 
\begin{eqnarray}
 R_s\equiv \frac{\int_0^\infty d\rho \rho^2 s'(\rho)^2}{\int
  _0^\infty d\rho \rho s'(\rho)^2}. \label{ini}
\end{eqnarray}
We also introduce the dimensionless temperature as
\begin{equation}
k\equiv \sqrt{\frac{2}{\epsilon^2}} \frac{g T}{4\sqrt{\kappa}\mu}= \sqrt{\frac{2}{\epsilon^2}} {T\over T_{\infty}},
\label{eq:defk}
\end{equation}
where as is shown in \eqref{JB}, the parameter for the thermal potential is written as
\begin{eqnarray}
{1\over 4\pi^2} \frac{m_B^2}{T^2}=\frac{g^4}{32\pi^2 \epsilon^2 \kappa^2 k^2} s^2<\frac{g^4}{32\pi^2 \epsilon^2 \kappa^2 k^2}.
\end{eqnarray}

\begin{figure}[htbp]
\begin{center}
 \includegraphics[width=.4\linewidth]{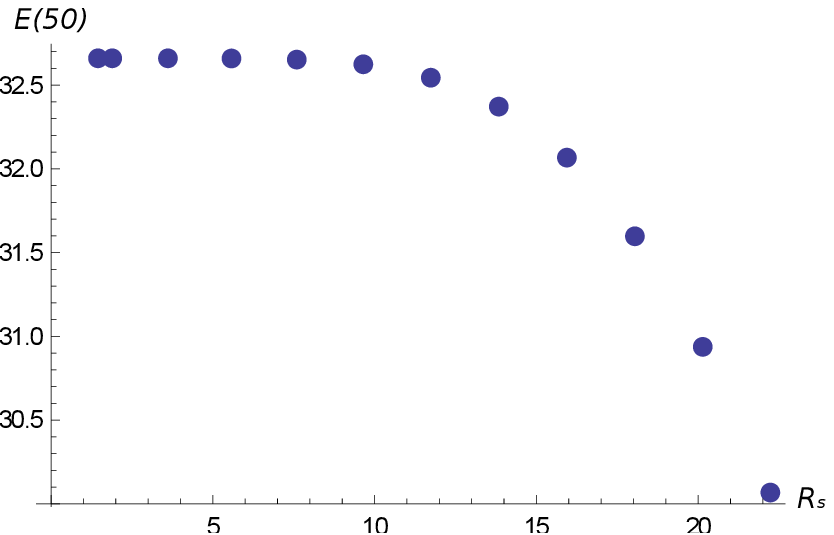}
 \includegraphics[width=.4\linewidth]{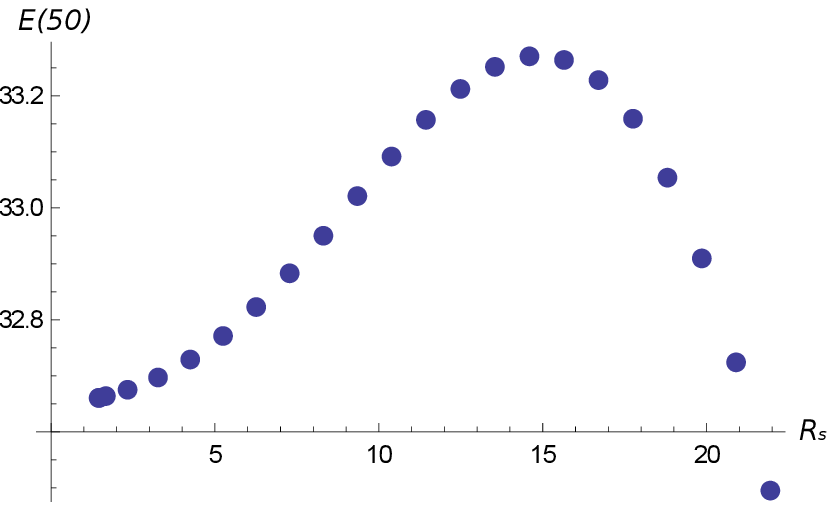}
\vspace{-.1cm}
\caption{\sl The low energy effective theory for $\lambda=0.27$,
  $\epsilon=1$ and $n=1$. Left panel is for almost zero temperature $k=0.001$. Right panel is for $k=0.01$. }
\label{EffectivePot1}
\end{center}
\end{figure}

To see how the stabilization mechanism works, we define the dimensionless energy as
\begin{eqnarray}
 E(\tau)\equiv \frac{\lambda}{m^2} \frac{E-E_{\rm vev}}{2\pi \Delta z}
=\frac{\lambda}{m^2} \frac{E}{2\pi \Delta z}-\frac12 {\cal V}(1)
\Lambda^2,  \label{eq:dimlessenergy}
\end{eqnarray}
at the relaxation time $\tau$ (for quick introduction to the relaxation method, see appendix A in \cite{our1}). 
Here we removed a contribution of the vacuum energy
density $E_{\rm vev }$. $\Lambda$ denotes the cut-off of the
energy and we set $\Lambda = 50$.
Since we cannot follow the evolution of the system in the relaxation method to $\tau\rightarrow \infty$, 
we evaluate it at $\tau=50$. 
In Figure \ref{EffectivePot1}, we plot $R_s$ and $E(\tau=50)$ for various initial conditions 
for $k=0.001$ and $k=0.01$. Here we take the parameters as   $\lambda=0.27$, $\epsilon=1$, and $n=1$.  
For these parameters, we can expect the R-string is stabilized for $k>\sqrt{2}T_1/T_{\infty}\approx 0.00155$. As one can see from the Figure, $R_s=0$ is the local energy minimum of the 
field configuration, 
which implies the R-string is stabilized for $k=0.01$, 
whereas it does not seem to be local minimum for $k=0.001$, as expected. 
Thus, we conclude that the R-strings are stabilized for $T>T_n$. 
It is interesting that the R-strings are stabilized even if the thermal mass for the messenger field $\sim g T$ 
is smaller than its zero temperature mass at the origin, 
$|m_\phi|=\sqrt{\kappa}\mu$. 
Finally, it is worth emphasizing that in the Figure \ref{EffectivePot1}, we draw the effective potential until relatively larger value of $R_s$ just for reference where high temperature expansion is no longer valid. Thus, only the behavior in small $R_s$ region is reliable, which is enough to guarantee the stability of the R-string. 

\subsection{Uplifted SUSY vacuum and stable R-tube}

We here point out that there is a phenomenologically safe situation even when the R-strings are 
unstable and they deform to R-tubes. One of the interesting features of thermal effects is lifting the lower energy
vacuum. Since SUSY is broken by the thermal effect, the SUSY vacuum can
be lifted and 
the SUSY-breaking vacuum becomes thermodynamically favored compared to the lower energy vacuum. 
To see that, let us focus on the messenger direction at $X=0$
and $|\phi|=|\tilde \phi|$. 
\begin{eqnarray}
 V_{\rm mgr}=\left(\frac{\kappa}{g^2}\hat \phi^2-\mu^2\right)^2
+T^4{\cal J}\left(\frac{\hat \phi^2}{ T^2}\right) ,
\end{eqnarray}
where $\hat \phi \equiv g |\phi|= g|\tilde \phi|$ and 
\begin{eqnarray}
 {\cal J}(x)=\frac{3}{2\pi^2} (J_B(x)-J_B(0)).
 \end{eqnarray}
Here, we subtracted $J_B(0)$ because the SUSY breaking vacuum also gets
lifted by thermal effect. It is given by constant term at
$\phi=\tilde{\phi}=0$. 
Thus, it is useful to subtract it and define the potential as above.
Here we can show the following inequalities 
\begin{eqnarray}
 0<x {\cal J}'(x)<{\cal J}(x).\label{eq:ineqJ}
\end{eqnarray}
The would-be SUSY vacuum is determined by
\begin{eqnarray}
 \frac{\partial V_{\rm mgr}}{\partial \hat \phi^2}=
-2 \frac{\kappa}{g^2}\left(\mu^2-\frac{\kappa}{g^2}\hat
		      \phi^2\right)+T^2{\cal J}'=0, \quad \Rightarrow \quad
 \hat \phi =\hat \phi_{\rm vac}\left(\frac{\kappa}{g^2},\frac{T}{\mu}\right)<
\frac{g\mu}{\sqrt{\kappa}} .\label{eq:upliftedvac}
\end{eqnarray}
The critical temperature, $T_{cr}$, at which the would-be SUSY vacuum and SUSY breaking vacuum become the same energy density, is determined by the following equation\footnote{Note that 
the field configuration in the realistic expanding Universe does not follow perfectly the R-tube solution. However, at least 
the R-tube solution is stable, the dangerous roll-over process will not cause and hence this 
consideration gives a good condition to avoid the roll-over problem. },
\begin{eqnarray}
 V_{\rm mgr}\Big|_{\hat \phi=\hat \phi_{\rm vac}}
=\mu^4\left(1-\frac1{4\lambda}\right),\quad \Rightarrow 
\quad   T= T_{cr}\left(\frac{\kappa}{g^2}, \lambda \right).
\end{eqnarray}
By taking derivatives with respect to $\kappa$ and $\lambda$ of the above
equation, we find that  
$T_{cr}(\kappa/g^2,\lambda)$ is a monotonically increasing function of 
$\kappa$ and $\lambda$ as 
\begin{eqnarray}
 \frac {\partial T_{cr}}{\partial \lambda}=\frac{
  \mu^4}{4\lambda^2}\left(\frac{\partial V_{\rm mgr}}{\partial
		     T}\right)^{-1}\Big|_{T=T_{cr},\hat \phi=\hat \phi_{\rm
  vac}} >0,\nn
g^2 \frac{\partial T_{cr}}{\partial \kappa}= 
2 \hat \phi^2 \left(\mu^2-\frac{\kappa}{g^2}\hat \phi^2\right)\left(\frac{\partial V_{\rm mgr}}{\partial
		     T}\right)^{-1}\Big|_{T=T_{cr},\hat \phi=\hat \phi_{\rm
  vac}} >0,\label{22siki}
\end{eqnarray}
where we used the inequalities (\ref{eq:ineqJ}). 

For $\lambda\gg 1$ or $\kappa\gg 1$, that is $\hat \phi_{\rm
vac}^2\ll 4\pi^2 T_{ cr}^2$, the potential reduces to
\begin{eqnarray}
 V_{\rm mgr}\approx \left(\frac{\kappa}{g^2}\hat
	      \phi^2-\mu^2\right)^2+\frac1{8} T^2\hat \phi^2 ,
\end{eqnarray}
and thus, the functions $\hat \phi_{\rm vac}$ and $T_{cr}$ are explicitly given as
 \begin{eqnarray}
 \hat \phi^2_{\rm vac}=\frac{g^2}{\kappa}\left(\mu^2-\frac{g^2
					  T_{cr}^2}{16\kappa}\right)
=\frac{g^2\mu^2}{2\kappa \sqrt{\lambda}},\quad
T_{cr}=\frac{4\sqrt{\kappa}}g\mu\times  \sqrt{1-\frac1{2\sqrt{\lambda}}}. 
\label{eq:highapprox}
 \end{eqnarray}
Then, 
the inequality (\ref{eq:funnyineq}) means that, for sufficiently large $\kappa$,   
\begin{eqnarray}
 T_1<T_{cr}<T_2.
\end{eqnarray}
In the opposite limit, $\lambda\sim 1/4$ or $\kappa \ll 1$, we also
find the explicit form\footnote{In our model, the gauge group is completely broken in the 
SUSY-vacuum and hence thermal logarithmic potential \cite{Anisimov:2000wx} for low temperature regime would not arise.
However, even if it appears, it also has an effect to lift up the SUSY-vacuum, and hence 
the basic feature does not change if it is positive. In the case of negative thermal logarithmic potential, the stabilization mechanism for R-tube does not work 
with  $\lambda\sim  1/4$ or $\kappa\ll  1$. } 
\begin{eqnarray}
 V_{\rm mgr}=\left(\mu^2 -\frac{\kappa}{g^2}\hat \phi^2 \right)^2+{\pi^2
  \over 30}T^4 \quad \Rightarrow \quad
  T_{cr}= \mu
\left(\frac{30}{\pi^2 }  \left(1-\frac1{4\lambda}\right)\right)^\frac14.
 \end{eqnarray}
For $\sqrt{\kappa}/g \ll 1$, therefore, we find the inequality $T_n < T_{cr}$ with
an arbitrary $n$. 

\begin{figure}[htbp]
\begin{center}
 \includegraphics[width=.41\linewidth]{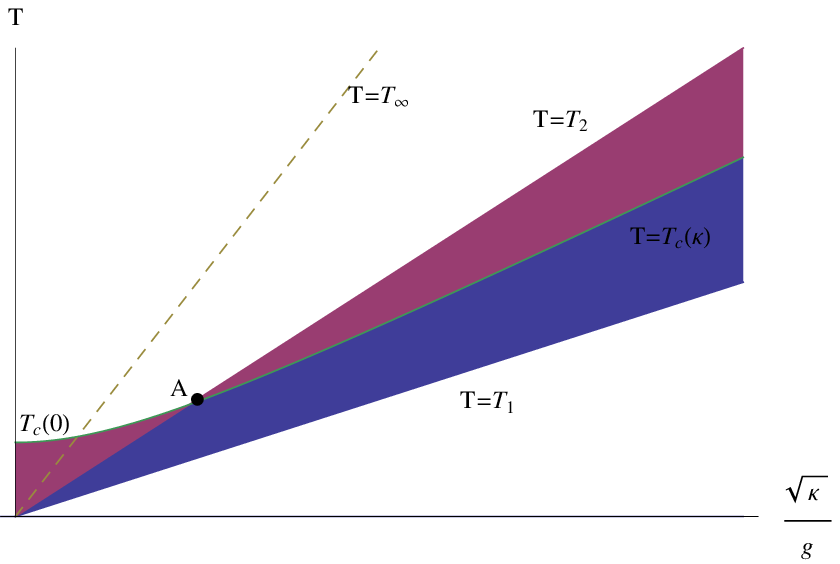}
 \qquad
  \includegraphics[width=.41\linewidth]{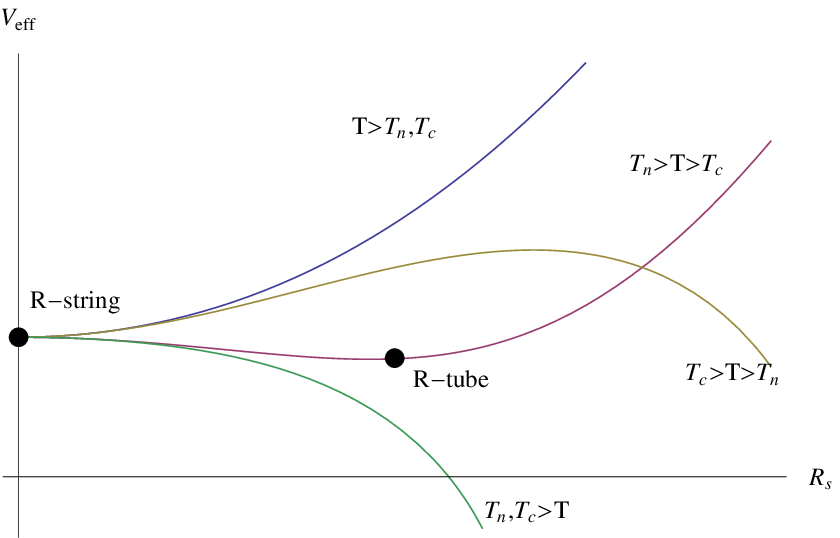}
\vspace{-.1cm}
\caption{\sl The left panel shows phases of R-string/R-tube.  
With a winding number $n$, an R-string is stable even including quantum
 effect at high temperature $T>T_n,T_{cr}$,
and is unstable in any sense at $T<T_{cr},T_n$. For $T_{cr}>T>T_n$, an
 R-string is classically stable and for $T_n>T>T_{cr}$ an R-tube is stable.  }
\label{fig:PhaseDiagram}
\end{center}
\end{figure}

Here, we show that the critical temperature of the R-string with
the minimal winding number is always lower than  $T_{cr}$. By using
(\ref{eq:upliftedvac}), \eqref{22siki} and \eqref{eq:highapprox}, we fine the following relation
\begin{eqnarray}
\frac{\partial T_{cr}}{\partial \kappa}<\frac{T_{cr}}{2\kappa}
\quad \Rightarrow \quad T_{cr}>\frac{4\sqrt{\kappa}}g\mu\times
\sqrt{1-\frac1{2\sqrt{\lambda}}}\, \quad (>T_1).  \label{TcT1}
\end{eqnarray}
{}From the properties noted in the above we can draw a phase diagram as
Figure \ref{fig:PhaseDiagram}. 
Note that an R-tube with the minimal winding number $n=1$ is always unstable.

\begin{figure}[htbp]
\begin{center}
 \includegraphics[width=.45\linewidth]{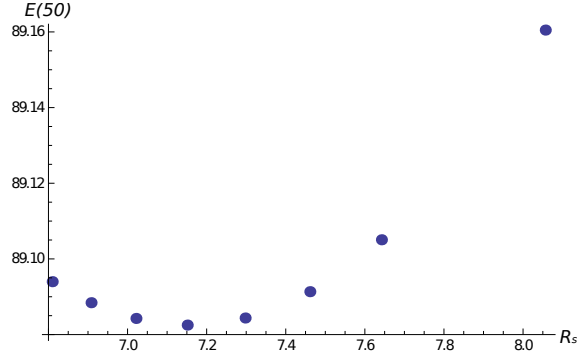}
\vspace{-.1cm}
\caption{\sl R-tube solution with $n=2$, $k=0.5$, $\lambda=0.27$,
  $\epsilon=1$. A minimum exists around $R_s\sim 7.2$. }
\label{n=2Tube}
\end{center}
\end{figure}

As an illustration, we show a stable R-tube solution with the winding
number $n=2$. We take parameters $\kappa=\epsilon=1$ and $k=0.5$, which allow us to use $\kappa \gg 1$ limit as (\ref{eq:highapprox}) though it is not good
enough. 
From the Figure \ref{TachyonMass1}, we see
that the tachyonic mass is $|m_2|\simeq (1.03/\sqrt{2})\times
\sqrt{\kappa}\mu$ for $\lambda=0.27$ and $\epsilon=1$. 
Since $k=0.5<1.03$, the R-string with $n=2$ is unstable and turns out
to become the R-tube in that
case. On the other hand, $k=0.5$ is enough to uplift the SUSY vacuum to
stabilize the R-tube as $k=0.5 > \sqrt{2-1/\sqrt{\lambda}}=0.275$. In the Figure \ref{n=2Tube} we show the low energy effective potential for the R-tube solution with $k=0.5$, $\lambda=0.27$, $\epsilon=1$. Clearly there exists the minimum around $R_s\sim 7.2$. 

Finally, let us comment on the validity to use the thermal
potential. In deriving the thermal potential, we assumed homogeneity
at least of order ${\cal O}(T^{-1})$. On the other hand, when we study
the stability of solution we put the R-string/R-tube solution which
generates inhomogenioety in the space. A condition of the validity to the thermal potential can be roughly estimated as 
\begin{equation}
{\partial \phi \over \phi }\leq T \iff  {\sqrt{\lambda }\mu^2 \over m}\tilde{\partial} s \leq T.
\end{equation}
In the stabilized R-string solution, there is no space dependence of
$s$. Thus this condition is satisfied. However, as for the R-tube
solution shown above, there is space dependence. 
Using the dimensionless temperature in (\ref{eq:defk}), 
we can estimate the validity condition,
\begin{equation}
\tilde{\partial} s \leq {4 \over g }k.
\end{equation}
The numerical analysis on the R-tube solution for $k=0.5$ leads to 
$\tilde{\partial} s\simeq 1/4$.
Thus, our numerical results are still under control 
for a proper value of $g$.

\section{Application to the expanding Universe}\label{sec:Universe}

Now we study how the discussion in the previous section is applied to the constraints 
on the realistic cosmology with expanding Universe.  Here we focus on the vacuum selection and the way to avoid the roll-over problem. 
There are also constraints from the cosmological gravitino or moduli problem, and R-axion particle abundances,  
but these are beyond the scope of the present paper. 
For the discussion on such problems, see \cite{FKT} and \cite{OurII}.

Let us follow cosmic histories in our scenario schematically. The thermal mass for the messengers is given by $m_{\phi}^T=c_\phi gT$, with $c_\phi$ being the numerical factor of the order of unity that counts the number of fields 
contributing the thermal mass. 
If messengers enter thermal plasma, it generates thermal mass to the $X$ field, as $m_X^T=c_X \kappa T$, 
where $c_X$ is the numerical factor of the order of unity. 
Note that during the inflaton oscillation dominated era, the temperature
and the Hubble parameter are related by 
\begin{equation}
T=\left({2\over 5} \right)^{1/4}\left({90 \over \pi^2 g_*
  }\right)^{1/8}
(T_R^2 M_{\rm pl} H)^{1/4}, 
\end{equation}
where $g_*$ is 
the effective relativistic degrees of freedom and $T_R$ denotes 
the reheating temperature. 
There are also the so-called Hubble induced mass for $X$ and messengers, 
$m_\phi^H \simeq m_X^H\simeq H$, 
which is generated from the Planck suppressed interaction between inflaton and these fields in supergravity. 
For small enough $\kappa$, 
\begin{equation}
\kappa< \left( {5\over 2}\right)^{1/4} \left(\pi^2 g_*  \over 90
\right)^{1/8} \left({|m_X|^3 \over T_R^2 M_{\rm pl} }\right)^{1/4} ,
\end{equation}
the Hubble induced mass overwhelms the thermal mass for $X$ fields when 
$H\gtrsim |m_X|$, and in the following, we consider such a situation. 
Here, we assume the inflaton oscillation dominated era since the gravitino problem requires relatively small reheating temperature. 

\begin{figure}[htbp]
\begin{center}
 \includegraphics[width=.4\linewidth]{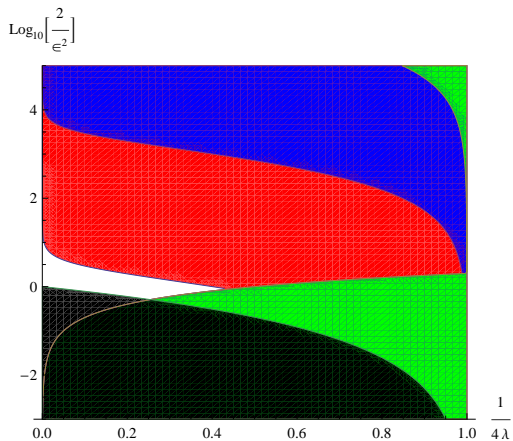}
\vspace{-.1cm}
\caption{The region satisfying one of the two inequality $T_R^{\rm gr}>T_A$ (defined in \eqref{upperbound}) and
 $|m_X|>|m_\phi|$ with
$(m_{3/2}/{\rm GeV}) \kappa^{-1/3}  = 10^{-6}, 10^{-3},  10^0$ which are
 colored by red, blue, and green respectively. The black region is excluded
 by the condition of the existence of the SUSY breaking vacuum.}
\label{Fig:vacuumselection}
\end{center}
\end{figure}

At a high temperature in the inflaton oscillation dominated era, the thermal masses or the Hubble induced mass
for $X$ and messenger fields restore the symmetry and 
these fields are set at the origin. 
As the temperature and the Hubble parameter decreases, 
the Hubble induced mass or the thermal mass can no longer fix the fields at the origin 
and symmetries are spontaneously broken. 
If the Hubble induced mass for the $X$ becomes inefficient earlier
than the thermal mass for messenger fields, 
the SUSY-breaking vacuum is naturally selected associated with the R-string formation. 
This condition can be written as
\begin{equation}
|m_\phi|=\sqrt{\kappa }\mu \, < \,m_{\phi}^T(H=m_X)=c_\phi g   \left({2\over 5} \right)^{1/4}\left({90 \over \pi^2 g_* }\right)^{1/8}(T_R^2 M_{\rm pl} m_X)^{1/4},  \label{cond1}
\end{equation}
which is rewritten in terms of the constraint on the reheating temperature,
\begin{equation}
T_R> T_A\equiv \left( { 5\pi^2 g_*  \over 72} \right)^{1/4} \left( {\kappa^2 \mu^4\over  (c_\phi g)^4  M_{\rm pl} m_X } \right)^{1\over 2}.\label{eq:recond1}
\end{equation}
In the other case, if $|m_X|>|m_\phi|$, 
the Hubble induced mass for $X$ always becomes
inefficient before that for $\phi,\tilde \phi$ become inefficient as
discussed in section \ref{sec:setup}. Therefore, one of the inequality
(\ref{redline}) and the inequality (\ref{eq:recond1}) is needed 
to be satisfied for selecting the SUSY-breaking vacuum as showed in the
Figure \ref{Fig:vacuumselection}.

After $X$ acquires the nonvanishing field value, the R-strings are formed and our discussion in the 
previous section can be applied. We have shown that any R-strings with arbitrary winding numbers 
are stabilized for sufficiently high temperature, 
\begin{equation}
T>T_{\infty}. 
\end{equation}
However, practically, it may be sufficient to protect only the mode
with $n=1$. 
In this case, the condition for the stabilization can be written as 
\begin{equation}
T>{\rm min}(T_1, T_{cr})=T_1. 
\end{equation}
As shown in \eqref{TcT1}, $T_{cr}$ is always higher than $T_1$. Since the R-string networks turn to the R-string-domain wall networks and immediately decay to R-axion particles 
at $H=m_a$ \cite{Sikivie:1982qv}, 
we can avoid the roll-over problem if the stabilization mechanism discussed above works at that time. 
As shown in the previous section, for $\sqrt{\kappa}/g \ll 1$, $T_n<T_{cr}$ is satisfied. 
Thus, the above constraints can be rewritten as follows, 

\begin{equation}
T(H=m_a)>T_\infty \Leftrightarrow \sqrt{\kappa }\mu < m_{\phi}^T(H=m_a)=
 c_\phi g \left({2\over 5} \right)^{1/4}\left({90 \over \pi^2 g_*
	   }\right)^{1/8} (T_R^2 M_{\rm pl} m_a)^{1/4}, 
\end{equation}
which is rewritten in terms of the constraint on the reheating temperature, 
\begin{equation}
T_R> T_B\equiv \left( { 5\pi^2 g_*  \over 72} \right)^{1/4} \left(
		{\kappa^2 \mu^4\over  (c_{\phi}g)^4  M_{\rm pl} m_a}
							    \right)^{1\over
2}=\sqrt{\frac{|m_X|}{m_a}}T_A, 
\label{condition2}
\end{equation}
and 
\begin{equation}
T(H=m_a)>T_1 \Leftrightarrow 
\frac{\epsilon |q_1|}{\sqrt 2}
\sqrt{\kappa} \mu 
<m_\phi^T(H=m_a))=c_\phi g \left({2\over 5} \right)^{1/4}\left({90 \over \pi^2 g_* }\right)^{1/8} (T_R^2 M_{\rm pl} m_a)^{1/4}, 
\end{equation}
which, in turn, is expressed as the constraint on the reheating temperature, 
\begin{equation}
T_R>T_C\equiv \left( {5\pi^2 g_* \over 1152} \right)^{1/4} \left(
	       {\epsilon^4q_1^4 \kappa^2 \mu^4 \over (c_\phi g)^4 M_{\rm
	       pl} m_a} \right)^{1/2}
=\left|\frac{\epsilon^2q_1^2}2 \right| T_B. 
\end{equation}
Note that a quantity  $|\epsilon^2q_1^2/2|^4(<1)$, which is
compared with a value in Eq.(\ref{eq:ratio}), can take 
a value $1$ in the limit of $\lambda\to \infty$ and also 
$10^{-22}$ in a case of $\lambda=0.27,\epsilon=1$, 
and thus there are both cases of $T_A>T_C$ and $T_A<T_C$
whereas inequalities $T_B> T_A,T_C$ are always satisfied. 

\begin{figure}[htbp]
\begin{center}
 \includegraphics[width=.55\linewidth]{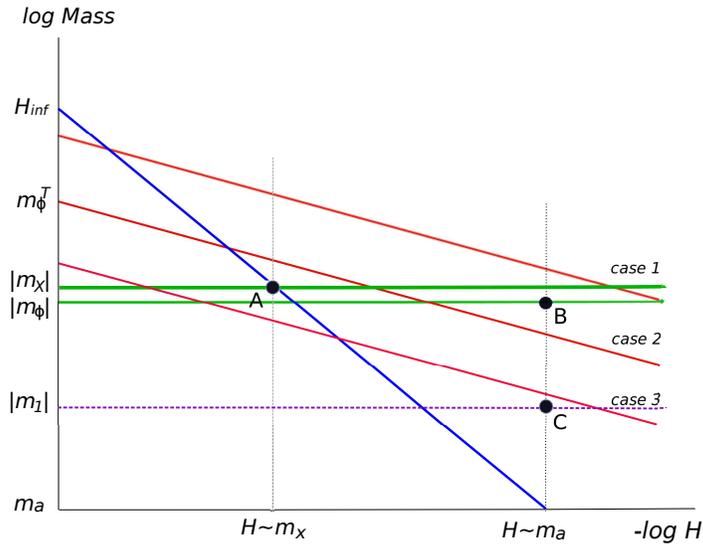}
\vspace{-.1cm}
\caption{\sl In the case 1, the reheating temperature is in $T_R> T_B$
  and the SUSY breaking vacuum is preferable. R-strings with all
  winding number are stable until their decay. In the case 2, the
  reheating temperature satisfies $T_B>T_R> T_A>T_C$. The vacuum selection is
  successful but some of higher winding R-strings are unstable. In the
  case 3, reheating temperature is lower $T_A>T_R>T_C$. The vacuum
  selection is unsuccessful when the tachyonic masses at the origin
  are $|m_{\phi}| > |m_X|$. }
\label{Fig:selection}
\end{center}
\end{figure}

In Figure \ref{Fig:selection}, we illustrate thermal histories by
showing Hubble parameter dependences of various mass terms. In this figure we assumed $T_A>T_C, |m_X|>|m_\phi|$ just to draw this figure concretely. The blue line is the Hubble induced mass linear in $H$. The red lines represent the thermal masses $m_{\phi}^T$ for the messengers with three different reheating temperatures. The horizontal green lines are the sizes of the tachyonic masses $|m_X|$ and $|m_\phi|$ at the origin with the following relation,
\begin{equation}
{|m_X|\over |m_\phi|}= {\epsilon \over 2 \sqrt{\lambda}}.
\end{equation}
Note that depending on the two parameters, the order of the two masses
is changed. Below we will show that the cases 1 and 2 in Figure \ref{Fig:selection}
realize the successful vacuum selection. 
At the point A, the Hubble parameter becomes comparable to the size of tachyonic mass of $X$ at the origin, so the vacuum starts to slide down to the SUSY breaking vacuum. 
Sine the condition \eqref{cond1} is satisfied in these two cases, the SUSY breaking vacuum is successfully chosen. 
The points B and C represent the points where $T(H=m_a) = T_\infty$
and $T(H=m_a) = T_1$ 
are satisfied, respectively.
In summary, the case 1 passes both the vacuum selection and stabilization of all R-strings with arbitrary 
winding number, whereas the case 2 passes the vacuum selection but stabilizes only the R-string with $n=1$. 
The case 3 is ruled out by the vacuum selection if
$|m_\phi|>|m_X|$. 
With typical values of $\lambda,\epsilon$, 
the condition (\ref{condition2}) is too strong to be satisfied, and the
case 2 or the case 3 are preferred where a value of $\epsilon^2q_1^2/2$
plays a quite important role.

As a reference, we compare these results to the constraint on the reheating temperature from the gravitino problem \cite{gravitinoproblem}, 
\begin{equation}
T_R<T_R^{\rm gr} \equiv 6.4\times 10^8 \ m_{3/2}\, \left({m_{\lambda} \over 100 {\rm GeV} } \right)^{-2}, \label{upperbound}
\end{equation}
where $m_\lambda$ denotes the gaugino mass.
This condition comes from the constraint for the thermally produced gravitinos at the time of reheating 
not to overclose the Universe. 
Note that there may be late time entropy production from the moduli decay, which would relax the constraint on the 
reheating temperature, but this upper bound on the reheating temperature is a good reference value. 

\begin{figure}[htbp]
\begin{center}
 \includegraphics[width=.45\linewidth]{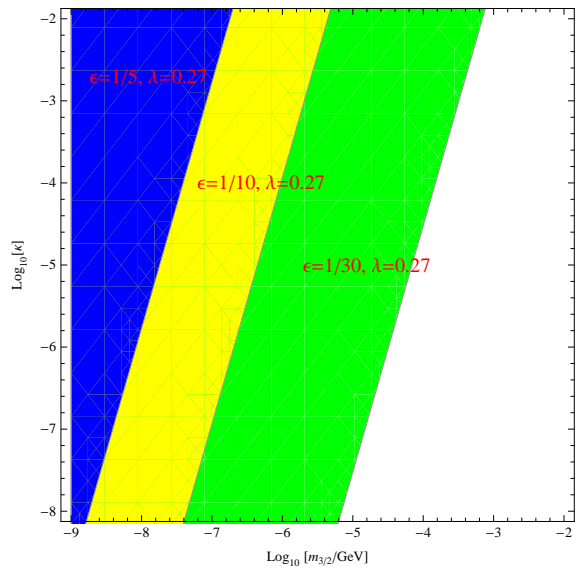}
 \qquad
\vspace{-.1cm}
\caption{\sl The regions satisfying the condition $T_B>T_R^{\rm gr}>
  T_A(>T_C)$ that correspond to to the case 2 in the Figure \ref{Fig:selection}. We take $g_*=220$ by assuming the messengers are not thermalized. 
 }
\label{AllowedTA}
\end{center}
\end{figure}

\begin{figure}[htbp]
\begin{center}
\includegraphics[width=.5\linewidth]{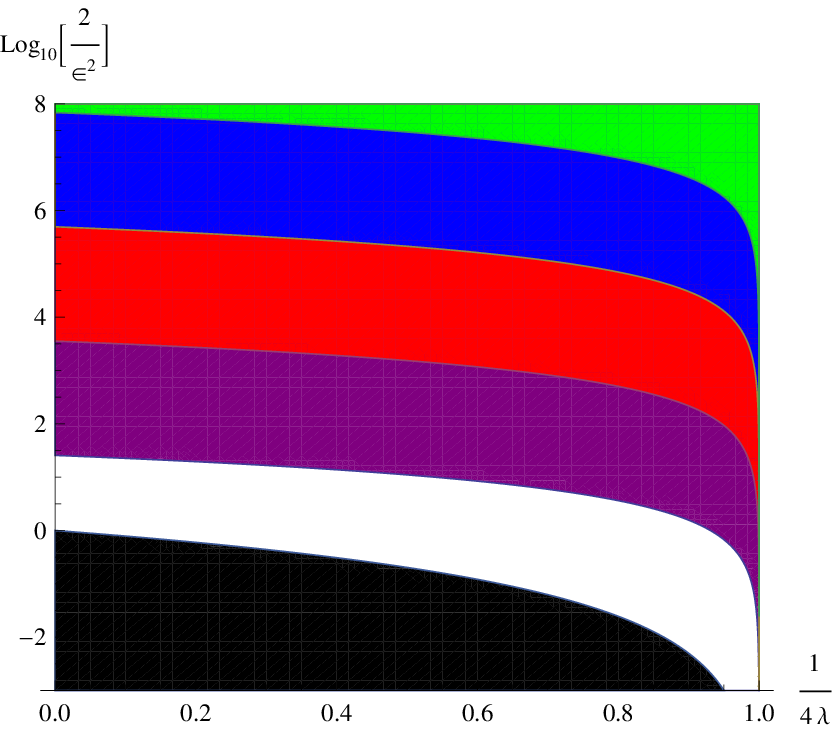} 
\includegraphics[width=.45\linewidth]{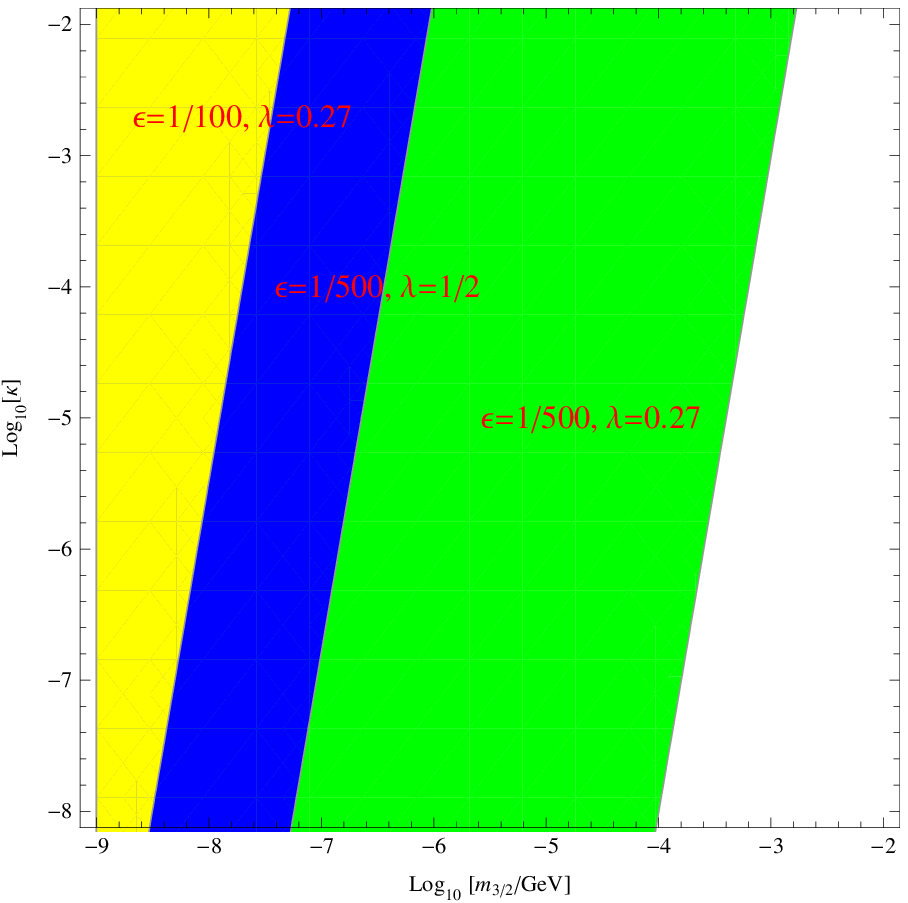}
\vspace{-.1cm}
\caption{\sl The regions corresponding to the case 1 in Figure
 \ref{Fig:selection},  $T_R^{\rm gr}> T_B$. 
 We assume $g_*= 220$. All winding strings are
  stable until the times of their decays. 
In the left panel the regions colored by 
purple, red, blue, and
 green  correspond to $(m_{3/2}/{\rm GeV}) \kappa^{-1/5}
  = 10^{-9}, 10^{-6}, 10^{-3},  10^0$, respectively. A region with
 $|m_X|>|m_\phi|$ is almost excluded by this strong condition.}
\label{AllowedTB}
\end{center}
\end{figure}

Figure \ref{AllowedTA} shows the window satisfying $T_R^{\rm gr}>T_A$ in the parameter space. 
We can see that there is a definite parameter space. 
For relatively large $\kappa$, a small gravitino mass is favored by the vacuum selection. 
Let us show the existence of non-zero window for $T_R^{\rm gr} > T_B$. In Figure \ref{AllowedTB}, we show regions satisfying $T_{R}^{\rm gr}> T_B$. 
Here the coefficient $c_\phi g$ is roughly estimated order one, $c_\phi g = {\cal O}(1)$. 
We adopt the gaugino mass in the gauge mediation as 
\begin{equation}
m_{\lambda}=  {\cal O}(0.1) \times {1 \over 16 \pi^2}
{g^{-1}(X_0)\sqrt{\lambda} \mu^2\over   \kappa m} ,
\end{equation}
where we assume a small hierarchy between gaugino mass and squark mass
by introducing the small parameter ${\cal O}(0.1)$ coefficient which can be interpreted as the R-breaking effect for Majorana gaugino mass. 
%
\begin{figure}[htbp]
\begin{center}
 \includegraphics[width=.45\linewidth]{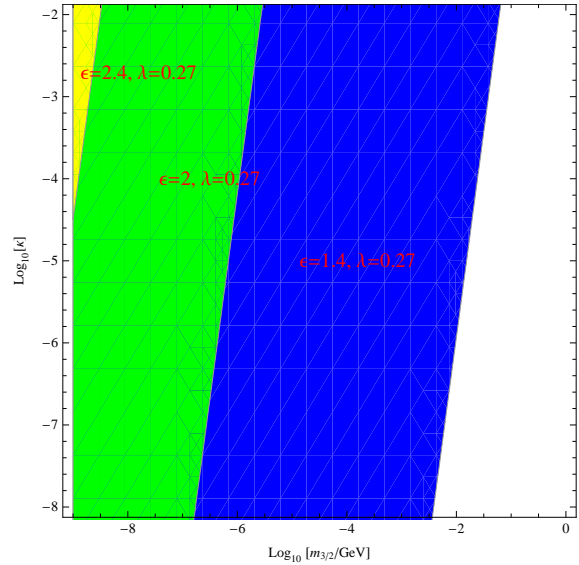}
\vspace{-.1cm}
\caption{\sl Three colored regions satisfy $T_A >T^{\rm gr}_R> T_C$ corresponding to the case 3 in Figure \ref{Fig:selection}. We took $\lambda=0.27$ and $g_*= 220$. Tachyonic masses we used $(q_{1}^2, \epsilon)=(10^{-5.4}, 1.4 ), (10^{-4.8}, 2 ), (10^{-4.3}, 2.4)$. }
\label{AllowedTC}
\end{center}
\end{figure}
%
Finally we show examples for the case $T_R^{\rm gr}>T_C$ in Figure ~\ref{AllowedTC}. 
Here we have used  the numerical results for the tachyonic masses shown in Figure \ref{TachyonMass1}. 

In conclusion, we surely have the scenarios that are free from the roll-over problem, 
avoiding the thermally produced gravitino problem. For this mechanism the value of a tachyonic mass of a fluctuation
around the R-string configuration, $q_1^2$, often plays a quite important
role. This value is very sensitive with details of the model
and can take a value with 
a quite wide range as we discussed, and thus,  
one has to always check this when building a more realistic model. Of course, there are also constraints from the moduli problem, the moduli induced gravitino problem, and R-axion problems, 
which are beyond the scope of the present study. 
We emphasize that our study is helpful for the future model building and studying the constraint on such models. 

\section{Discussion}

In this paper, we focused on gauge mediation models and studied the thermal potential to the messenger generated by the thermal plasma of the standard model particles. However, in the gravity mediations, interactions to the standard model particles are suppressed by the Plank mass. It is not obvious to get large enough thermal potential to stabilize the unstable modes. However, in this case, there is one interesting possibility to stabilize unstable modes. Supergravity corrections also provide a positive mass term to the fields $\phi$ and $\tilde{\phi}$, 
\begin{eqnarray}
V_{SUGRA}&=&{ |\phi|^2 \over M_{\rm pl}^2} \left( g^{-1}_{X\bar{X}}(0)-g^{-1}_{X\bar{X}}(X_0)  \right)|f|^2 + |\phi|^2 |c|^2+\cdots \\
&=& m_{3/2}^2 |\phi|^2 \left( \frac{3(g^{-1}_{X\bar{X}}(0)-g^{-1}_{X\bar{X}}(X_0))}{ g^{-1}_{X\bar{X}}(X_0)}+1 \right) +\cdots \\ 
&=&\frac{1+1/2 \lambda}{1-1/4\lambda}m_{3/2}^2  |\phi|^2  +\cdots ,
\end{eqnarray}
where $f=\mu^2$ is the F-term of $X$ and $c$ denote the constant term in the superpotential, 
we have not explicitly written similar terms for $\tilde{\phi}$
and we used the condition for the cancellation of the cosmological
constant. From the last equality, we see that the mass is positive. In
gravity mediation models, the SUSY breaking scale is relatively
large. Therefore, if the gravitino mass satisfies the following
inequality, 
\begin{equation}
|m_n|
 < m_{3/2}.
\end{equation}
then it may be possible to stabilize R-strings with low winding number. 
As long as $\kappa$ is very small, this condition is satisfied. Also, it is worth mentioning that in such a large SUSY breaking, the axion mass can be larger than the symmetry breaking scale, $m_a> |m_X|$. In this case, explicit breaking effects are significant, so the R-string is immediately broken after its formation.

From our study, it may be plausible to assume that in a realistic
gauge mediation model with spontaneous R-breaking, rolling over the
potential hill by the unstable solitons does not occur. In this case,
we have to impose further cosmological conditions, such as the moduli problem 
or the moduli induced gravitino problem studied in Ref.~\cite{FKT}.
Moreover, as intensively
studied in \cite{OurII}, when the R-string decay by axionic domain
wall, it produces a large amount of R-axions. As for the long-lived R-axion, cosmological constraints such as Big Bang Nucleosynthesis and CMB observation severely constrain the parameter space \cite{OurII}.

\section*{Acknowledgement}

The authors would like to thank W. Buchm\"uller, Y. Hamada, and T. Konstandin for useful comments and discussions. TK is supported in part by the Grant-in-Aid for the Global COE Program "The Next Generation of Physics, Spun from Universality and 
Emergence" and the JSPS Grant-in-Aid for Scientific Research
(A) No. 22244030 from the Ministry of Education, Culture,Sports, Science and 
Technology of Japan. YO's research is supported by The Hakubi Center for Advanced Research, Kyoto University.

\appendix

\section{Property of $a_n$\label{propertyan}}
\begin{figure}[htbp]
\begin{center}
 \includegraphics[width=.45\linewidth]{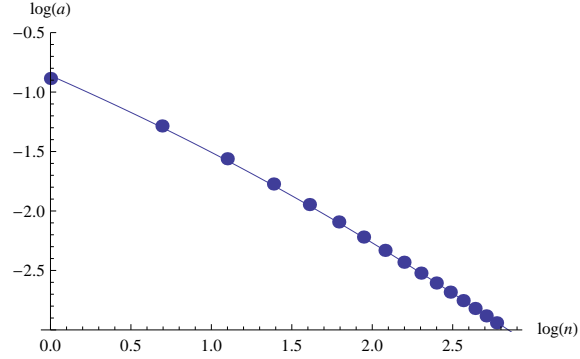}
\vspace{-.1cm}
\caption{\sl Dots are numerical data of $a=\tilde a_n$ and a solid line
 gives $a=1/\sqrt{4.5 n+1.1 n^2}$.}
\label{fig:Coefficient}
\end{center}
\end{figure}

Near the core of R-string, the solution behaves as 
\begin{eqnarray}
 A_n^{\rm sol}(\rho) =(a_n\rho)^n  \quad   {\rm for~}~ a_n \rho \ll 1.
\end{eqnarray}
For large $\lambda$ a property of a coefficient $a_n$ 
can be discussed as follows.  
Let us consider  the following limit of the action after extracting the
vacuum energy 
\begin{eqnarray}
 \lim_{\lambda \to \infty}\tilde S= -2\pi \int_0^\infty d\tilde \rho 
\tilde \rho
\left\{\left(\left(\frac{d A_\infty}{d
	      \tilde \rho}\right)^2+\frac{n^2}{\tilde \rho^2}A^2_\infty\right)+\frac1{4}(1-A^2_\infty)^2\right\}
+{\rm constant}, \label{effac}
\end{eqnarray}
with a redefinition of the function and  the coordinate
\begin{eqnarray}
 A_{\infty}(\tilde \rho)=\lim_{\lambda\to \infty} 
A\left(\rho=\sqrt{\lambda}\tilde \rho\right). 
\end{eqnarray}
In this limit, we also have a string solution $A_\infty(\tilde \rho)
=f_n(\tilde \rho)$ with
winding number $n$ and  this solution behaves as
\begin{eqnarray}
 A_\infty(\tilde \rho )=f_n(\tilde \rho ) \sim  (\tilde a_n \tilde
  \rho)^n   \quad {\rm for~}~  \tilde a_n \tilde \rho\ll 1.  
\end{eqnarray} 
Note that $\tilde a_n$ is a constant and depends on only the number $n$.
We know that the radius of R-string is estimated as $\tilde \rho \sim 2n$,
 and this fact implies
that 
\begin{eqnarray}
f_n(2n) ={\cal O}(1) \quad \Rightarrow\quad   
\tilde a_n = \frac1{n} \times  {\cal O}(1). 
\end{eqnarray}
Figure \ref{fig:Coefficient} shows the numerical result in the limit of $\lambda \rightarrow \infty$ \eqref{effac}, 
where ${\tilde a}_n$ is well approximated by ${\tilde a}_n =1/\sqrt{4.5n+1.1 n^2}$. 
 Therefore, we conclude that $\lambda$ dependence of  $a_n$ for large
 $\lambda$ (and sufficiently large $n$) is given by 
\begin{eqnarray}
 a_n = \frac{\tilde a_n}{\sqrt{\lambda}}=\frac{1}{n\sqrt{\lambda}}
  \times {\cal O}(1)  \quad {\rm for~}  \lambda \gg 1.
\end{eqnarray}

\section{Properties of the tachyonic mass}\label{sec:tachyon}
The first excited mode around the R-string has the point $\rho=\rho_c$ so
that
\begin{eqnarray}
\frac{d}{d\rho}\left(\rho \frac{d
		s}{d\rho}\right)\Big|_{\rho=\rho_c}=0,\quad
\Leftrightarrow \quad  V_{\rm pot}(A_n^{\rm sol}(\rho_c))=q_n^2.
\label{eq:tachyon1}
\end{eqnarray}
This value of $\rho_c$ must represent a scale of this excitation mode
and thus the following quantity should be of order one,
\begin{eqnarray}
 \gamma_n\equiv \sqrt{q_n^2-V_{\rm pot}(0)} \times \rho_c= {\cal O}(1).\label{eq:tachyon2}
\end{eqnarray} 
The above two equations give rough estimations of $q_n^2, \rho_c$ as
follows.
Let us assume that the following inequality 
\begin{eqnarray}
 1+\frac{\epsilon^2 q_n^2}{2}\ll \frac{2}{\epsilon^2}+\frac1{2\lambda}.\label{eq:ineq1}
\end{eqnarray}
Under this assumption (\ref{eq:tachyon1}) can be solved as
\begin{eqnarray}
\left(A_n^{\rm sol}(\rho_c)\right)^2\approx \left(\frac2{\epsilon^2}+\frac1{2\lambda}\right)^{-1}
  \left(1+\frac{\epsilon^2 q_n^2}{2}\right)\ll 1. 
\end{eqnarray}
Therefore  we can use $A_n^{\rm sol }(\rho) \approx (a_n \rho)^n$ under
the assumption
and combining (\ref{eq:tachyon2}) we find 
\begin{eqnarray}
A_n^{\rm sol}(\rho_c)\approx (a_n \rho_c)^n =
\frac{\epsilon^2 \gamma_n}2 \left(\rho_c \sqrt{1+\frac{\epsilon^2}{4\lambda}}\right)^{-1}.
\end{eqnarray}
This result reads
\begin{eqnarray}
a_n \rho_c&=&\left(\frac{\epsilon^2 a_n \gamma_n}{2\sqrt{1+\frac{\epsilon^2}{4\lambda}}}\right)^{\frac1{n+1}},\quad \nn
\frac{\epsilon^2 q_n^2}2&=&-1+\frac{\epsilon^2 \gamma_n^2}{2\rho_c^2}
=-1+\left(\frac{\epsilon^2}2\right)^{\frac{n-1}{n+1}}
\left(1+\frac{\epsilon^2}{4\lambda}\right)^{\frac1{n+1}} (a_n\gamma_n)^{\frac{2n}{n+1}}.
\end{eqnarray}
The assumption (\ref{eq:ineq1}) is satisfied if
\begin{eqnarray}
  \sqrt{1+\frac{\epsilon^2}{4\lambda}}\gg\frac{\epsilon^2a_n \gamma_n}2.
\end{eqnarray}
Let us remember that a value of $a_n^{-1}$ is proportional to $2n$ for
large $n$ and $\sqrt{\lambda}$ for large $\lambda$ and $\gamma_n$ can be expected to be
$\gamma_n\approx 1$ although $\gamma_n$ depends on $\lambda,\epsilon$
and $n$ slightly.
Sufficiently large $n$, large $\lambda$ and small $\epsilon$ satisfy the above and
therefore in several limits  we find
\begin{eqnarray}
\lim_{n\to \infty}\frac{\epsilon^2 q_n^2}2=-1,\quad  \lim_{\lambda\to \infty}\frac{\epsilon^2 q_n^2}2=-1,\quad 
 \lim_{\epsilon\to 0}\frac{\epsilon^2 q_{n>1}^2}2=-1,
\end{eqnarray}
whereas the case with $n=1$ gives an important exception as 
\begin{eqnarray}
 \lim_{\epsilon\to 0}\frac{\epsilon^2 q_{1}^2}2=-1+a_1
 \times  \gamma_1|_{\epsilon\to 0}.
\end{eqnarray}

\end{document}